\documentstyle[11pt,IAU207_pasp,twoside]{article}
\input epsf

\def\edcomment#1{\iffalse\marginpar{\raggedright\sl#1\/}\else\relax\fi}
\reversemarginpar

\begin{document}
\title{Globular Cluster Subpopulations in Early-Type Galaxies:\\
Insights into Galaxy Formation}
 \author{Jean P. Brodie}
\affil{University of California Observatories/Lick Observatory, University of California, Santa Cruz, CA 95064, U.S.A.}

\begin{abstract}
New and archival HST images of the globular cluster systems of 17
relatively nearby galaxies and 6 more distant galaxies have been used to
establish the characteristics of GC subpopulations over a range of host
galaxy morphological type from E5 to Sa.  GC color/metallicity, size and
luminosity distributions have been examined in the nearby galaxies and
color distributions have been determined for the more distant sample.
Correlations with parent galaxy properties and trends with galactocentric
radius have been explored.  Supplemented with Keck spectroscopy, our
results are best explained by an {\it in situ} formation scenario in which
{\it both} GC subpopulations formed at early times within the potential
well of the protogalaxy, in multiple episodes of star formation.  We have
also discovered a third population of clusters, fundamentally distinct from
the compact red and blue clusters common in early type galaxies.
\end{abstract}

\section{Introduction}

Since their discovery in early-type galaxies nearly a decade ago, globular
cluster (GC) subpopulations have been widely recognized as holding
important clues about both GC and host galaxy formation.  The presence of
subpopulations is apparent from the bimodal color
distributions in many early-type galaxies, implying multiple
mechanisms or epochs of GC formation in a galaxy's history. 

Explanations for bimodality fall into three categories: Mergers, {\it in
situ}/multi-phase collapse and accretion scenarios. The merger scenario,
in which elliptical galaxies
form from gaseous mergers of spiral galaxies
(Schweizer 1987; Ashman \&
Zepf 1992), was the only one to actually predict bimodality rather than
explaining it after the fact. The GC system of the
resulting galaxy consists of a blue (metal poor) population from the
progenitor spiral galaxies and a new population of red (metal rich)
clusters formed in the merger event itself out of enriched gas. The {\it in
situ}/multi-phase scenario was put forward by Forbes, Brodie \& Grillmair
(1997) in light of an increasing body of observational evidence not easily
explained under the merger model.  In this scenario the blue GCs form first
in a clumpy protogalactic medium. Star and cluster formation is halted
after most of the blue clusters and a few stars have formed, perhaps
because of the ejection of gas from these clumps due to supernova
explosions. There follows a dormant period of a few Gyr while the
gas expands, cools and falls back into the now more fully formed galaxy
potential. At this point star and cluster formation starts again, this time
from enriched gas, forming the red GCs and the bulk of the galaxy
stars. Harris et al.~(1999) came to similar conclusions based on
their HST study of GCs and halo stars in NGC 5128. In
accretion scenarios (e.g. C{\^o}t{\'e} et al 1998) the red GCs form
{\it in situ} in large elliptical galaxies and the blue GCs are
accreted along with lower luminosity galaxies or they can be stripped from
neighboring galaxies. Some new cluster formation may also be required
(Hilker et al.~1999).

Clearly, mergers take place and massive clusters are formed in the process.
This is frequently observed (e.g. Schweizer, 1997). Equally clearly, more
massive galaxies accrete less massive ones, along with their retinues of
more metal poor GCs.  These phenomena must influence the final
characteristics of a GC system, but to what extent? The questions we need
to address are: Is there a dominant mechanism determining the global
properties of GC systems in ellipticals? Is that mechanism the same for all
galaxies. How do spirals fit into this picture? If GC systems are the
result of a mix of processes how does that mix vary with host galaxy
properties? A point to remember, though, is that at high enough redshift
the distinction between the scenarios is quite blurred.

We have explored the red and blue GC subpopulations of 23 galaxies observed
with the WFPC2 camera on HST. Much of the data is from our own HST programs
but we have supplemented with data from the HST archive where these are of
sufficient depth. The result is a large, {\it homogeneous} set of GC
system data observed with the same telescope (HST) and instrument (WPFC2)
through the same filters (except for two galaxies) and all
subjected to the same reduction and analysis techniques. This homogeneity
is the key to uncovering subtle correlations which can be masked by
systematics when comparing data sets from the literature which have been
observed and analyzed in a variety of different ways.

Our approach has been to determine the key properties of the GCs (ages,
metallicities, sizes, luminosity function turnover magnitudes), relate
them to characteristics of the parent galaxy (luminosity, color,
environment) and try to identify trends with galactocentric radius. In
addition to the HST data we have used Keck spectroscopy to provide more
accurate age and abundance estimates for selected clusters. This
information has allowed us to put some useful constraints on the formation
history of GC systems and their host galaxies.

\section{Nearby Galaxies}
  
Our nearby sample comprises 17 galaxies out to the distance of the Virgo
cluster. It is discussed in detail in Larsen et al.~(2001) so I will provide
only a summary and a few updates here.  In Larsen, Forbes
\& Brodie (2001) we present additional information on the Sombrero (Sa)
galaxy. Included in our nearby sample are 1 Sa, 4 S0, 11 E galaxies and 1
cD galaxy. They cover a range in M$_B$ from $-$18.6 to $-$22.1 mag. Some are
found in groups, some in rich clusters and a few are in relatively isolated
environments.

\subsection{Color Magnitude diagrams}

The examples in Figure 1 illustrate the point that some color-magnitude
diagrams are obviously bimodal to the eye, while others are
not. Nonetheless, a KMM test (Ashman, Bird \& Zepf 1994) indicates, with
high probability, that two Gaussians are a better fit to the data than a
single one in nearly all cases. Interestingly, the peak $(V-I)_o$ colors
are consistent even in galaxies where the CMDs and $(V-I)_o$ histograms
show weak/no evidence for bimodality. One galaxy, NGC 4365, appears to have
only one (broadened) peak, at $(V-I)_o = 0.98$, typical for the metal poor
populations in bimodal systems.  This galaxy, a bright elliptical, would
be expected to possess a significant metal rich population under the
accretion scenario.

\begin{figure}
\begin{center}
\begin{minipage}{4in}
\epsfxsize=4in
\epsfbox{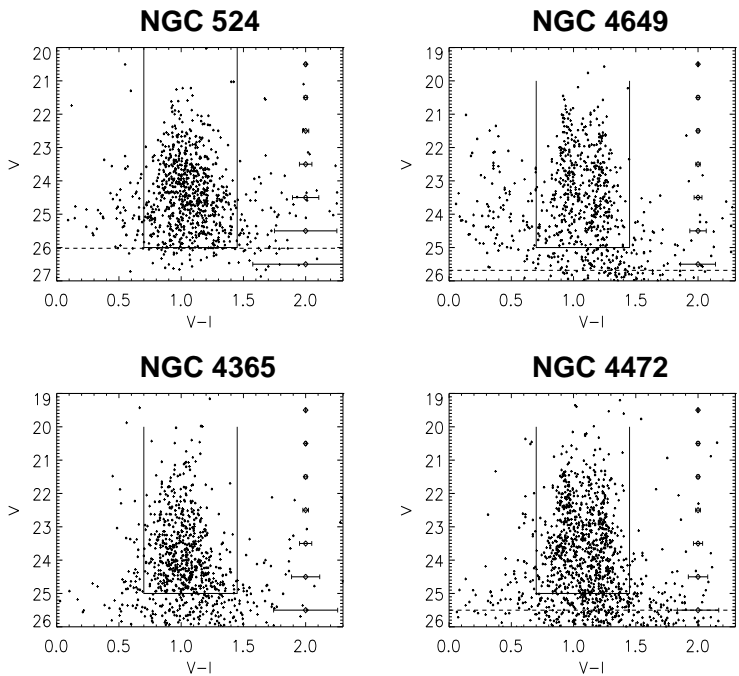}
\end{minipage}
\end{center}
\caption{Selected CMDs. The right panels are obviously bimodal while the
left panels are not. Only NGC 4365 is found to be unimodal by the KMM
test.}
\end{figure}

The average blue and red peak $(V-I)_o$ colors are 0.95$\pm$0.02 and
1.18$\pm$0.04, corresponding to $[Fe/H]=-1.4$ and $-0.6$ respectively
(Kissler-Patig et al.~1998). This is remarkably similar to the
metallicities of the Milky Way metal poor (halo) and metal rich
(disk/bulge) GCs (Zinn, 1985).

\subsection{Globular Cluster Luminosity Functions}

The GC luminosity function (GCLF) is usually assumed to have a Gaussian
form (although a students t$_5$ function is actually a better fit) whose
peak appears to occur at a constant absolute magnitude, modulo metallicity
effects (Ashman, Conti \& Zepf 1995).  The physical basis for the observed
GCLF is unknown as is the degree to which it is universal. Its shape may
have been set up at the time of formation or it may be the result of
dynamical evolution. For example, dynamical effects could have eroded the
faint end of a power-law initial luminosity function, resulting in the
GCLF we observe today.

Differences between the turnover magnitudes of the blue and red
subpopulations (${\Delta}m_V^{TO}$) would be indicative of differences in
their formation and/or subsequent evolution.  Assuming the same mass
function for the blue and the red GCs, similar old ages and metallicities
of [Fe/H] $= -$1.6 and $-0$.6, 1996 versions of the Bruzual \& Charlot SSP
models predict a difference in m$_V^{TO}$ between the blues and the reds of
0.26 mag.

We have examined the GCLF turnover magnitudes for the galaxies in our
nearby sample and find $<{\Delta}m_V^{TO}>=$0.47 mag., or 0.30$\pm$0.16
mag.~if we exclude systems for which the error on ${\Delta}m_V^{TO}$ is
$>$0.25. Note, though that this may introduce a bias because less populous
systems belonging to lower-luminosity galaxies tend to be excluded. The fact
that our observational result is close to the theoretical expectation
suggests that both populations are indeed old {\it if} their mass functions
are identical. However, if these populations formed at different epochs under
different environmental conditions, differences in their mass functions would not be altogether surprising.

\subsection{The Discovery of a Third Class of Cluster}

Observational limitations have meant that little has been known until now
about the faint wing of the GCLF, yet this is where the signatures of
formation and dynamical evolution should be the strongest.  At $m-M\sim$30,
the nearby S0 galaxy, NGC 1023 is close enough that the entire faint wing
of the GCLF is accessible for high S/N imaging with HST and
spectroscopy with 8--10m class telescopes.

\begin{figure}
\begin{minipage}{13.4cm}
\begin{minipage}{6.6cm}
\epsfxsize=6.6cm
\epsfbox[85 370 535 695]{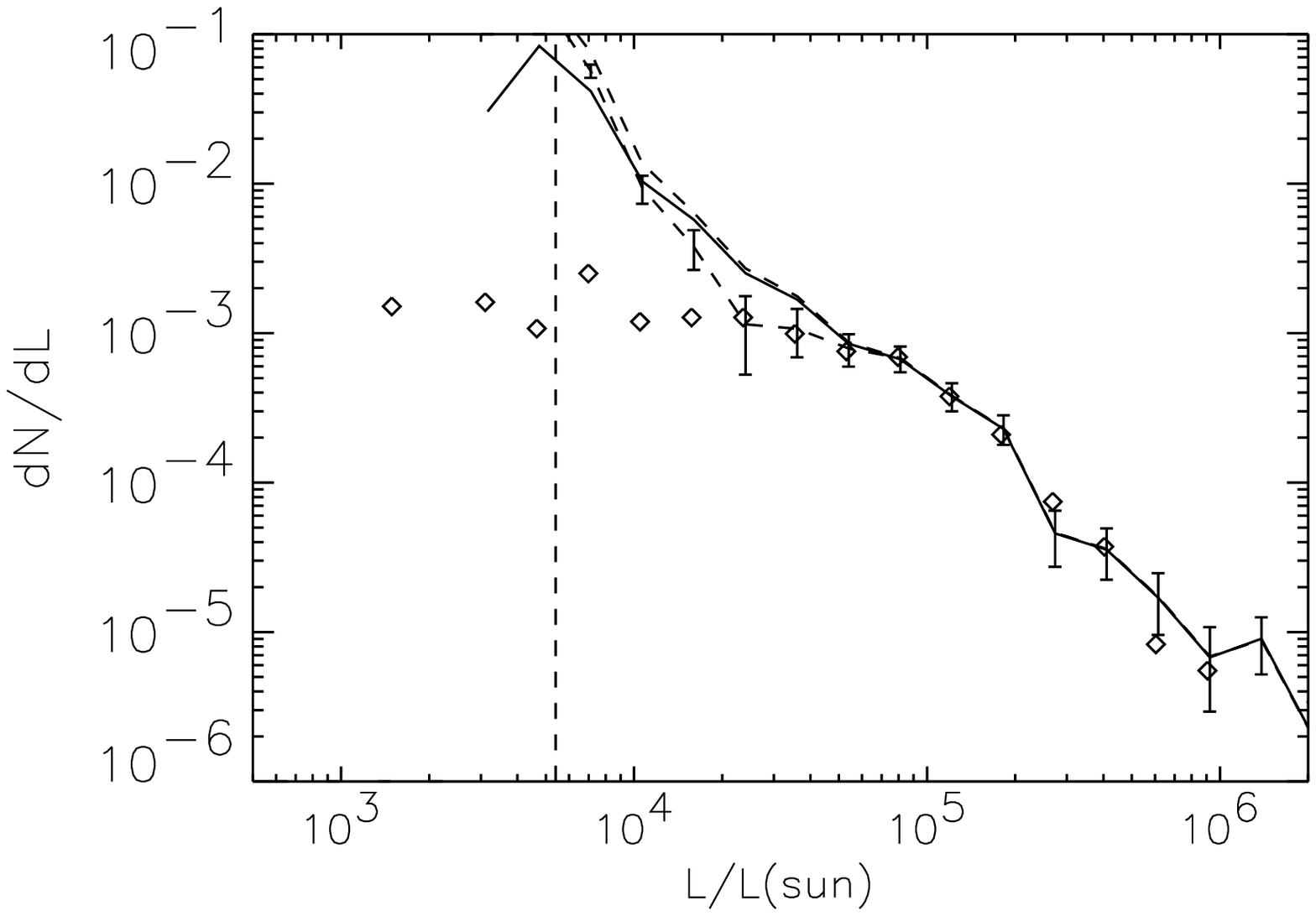}
\end{minipage}
\begin{minipage}{6.6cm}
\epsfxsize=6.6cm
\epsfbox[85 370 535 695]{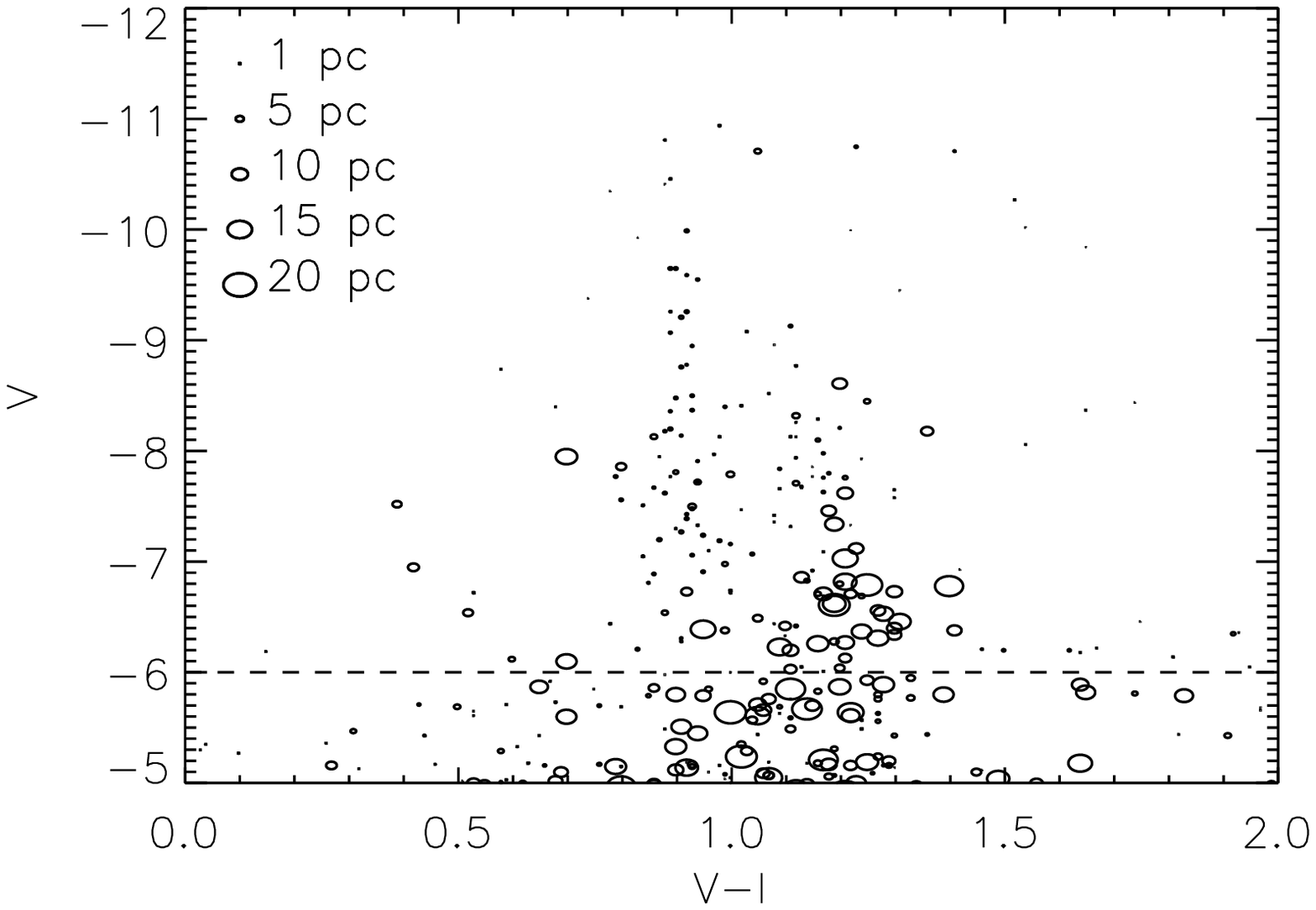}
\end{minipage}
\end{minipage}
\caption{Left panel: 2a) The GCLF of NGC 1023 (solid line) is
significantly different from that of the Milky Way (open circles) at low
luminosities. Right panel: 2b) A population of faint, red, extended objects is
apparent in this CMD where symbol size is proportional to object size.}
\end{figure}

As Figure 2a shows, the faint wing of the GCLF in NGC 1023 deviates
significantly from that of the Milky Way. This due to the presence of a
third population of clusters in addition to the normal compact blue and red
subpopulations commonly found in early type galaxies (Figure 2b). These
hitherto undetected faint, red extended objects are aligned with the galaxy
isophotes and thus appear to be associated with its disk. These ``faint
fuzzies'' are discussed in detail in Larsen \& Brodie (2000). They have no
analogs in the Milky Way and it is impossible at present to assess how
common they might be. They are obviously difficult to detect, being
extended low surface brightness objects fainter than the GCLF turnover and,
of the galaxies studied with HST in sufficient depth for accurate size
measurements, they would have been detectable in only 4 cases. They do
appear to be present in NGC 3384, another lenticular galaxy, but are
definitely absent in the populous lenticular NGC 3115.

In speculating about the origin of these objects we note that there is a
nearby dwarf companion galaxy containing two brighter blue objects for
which we have Keck spectra. These objects are young clusters (ages
$\sim$300 Myr) possibly formed as a result of the interaction with NGC
1023.  Perhaps the faint fuzzies are older remnants of similar interactions
with long-ago digested companion galaxies.

\subsection{Sizes}

Sizes differences between blue and red GC populations had been noticed
previously in four galaxies by ourselves (NGC 4472: Puzia et al.~1999; NGC
1023: Larsen \& Brodie 2000) and others (NGC 3115: Kundu \& Whitmore 1998;
M87: Kundu 1999). With our new large sample we could assess how
common this phenomenon might be. We found that the blue GCs are always
larger than the reds by about 20\%. Figure 3 gives examples. Although
suitable multiple pointings exist for only a few galaxies, the size
difference between the blue and red GCs persists at all radii in these
systems.  Interestingly, the same size difference is seen in the Milky Way,
and at all radii. Note again, the similarities between the early type
galaxies and the spirals (Milky Way and Sombrero).

\begin{figure}
\begin{center}
\begin{minipage}{85mm}
\epsfxsize=85mm
\epsfbox[35 515 483 720]{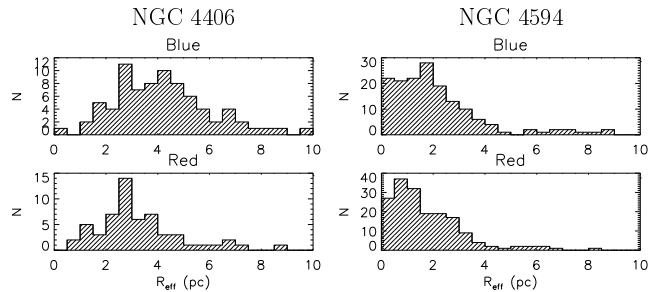}
\end{minipage}
\end{center}
\caption{GC size (R$_{eff}$) distributions. 
 Blue GCs (upper panels) are roughly 20\% larger than red GCs.}
\end{figure}

\subsection{Correlations with Parent Galaxy Properties}

Brodie \& Huchra (1991) showed that GC mean metallicity correlates with
parent galaxy luminosity. With the subsequent discovery of multiple GC
populations the question naturally arises as to whether the correlation
exists for one or both of the subpopulations separately.

In Figure 4 we show that the peak colors of {\it both} the red and the blue
GCs correlate with parent galaxy M$_B$ at the 2-3 ${\sigma}$ confidence
level. A similar relation exists for $(V-I)_{0}$ vs.~central velocity
dispersion. The slope of the relations is steeper for the red GCs than for
the blues. There is a 4 ${\sigma}$ correlation between the peak colors of
the red GCs and parent galaxy color\footnote{The lack of correlation
reported previously was due to the use of inaccurate galaxy colors. New
galaxy colors from Tonry et al.~(2001) are used here and reveal the
expected correlation.} and a 2 ${\sigma}$ correlation for the blues.

\begin{figure}
\begin{minipage}{14cm}
\begin{minipage}{6.9cm}
\epsfxsize=6.9cm
\epsfbox[85 370 535 695]{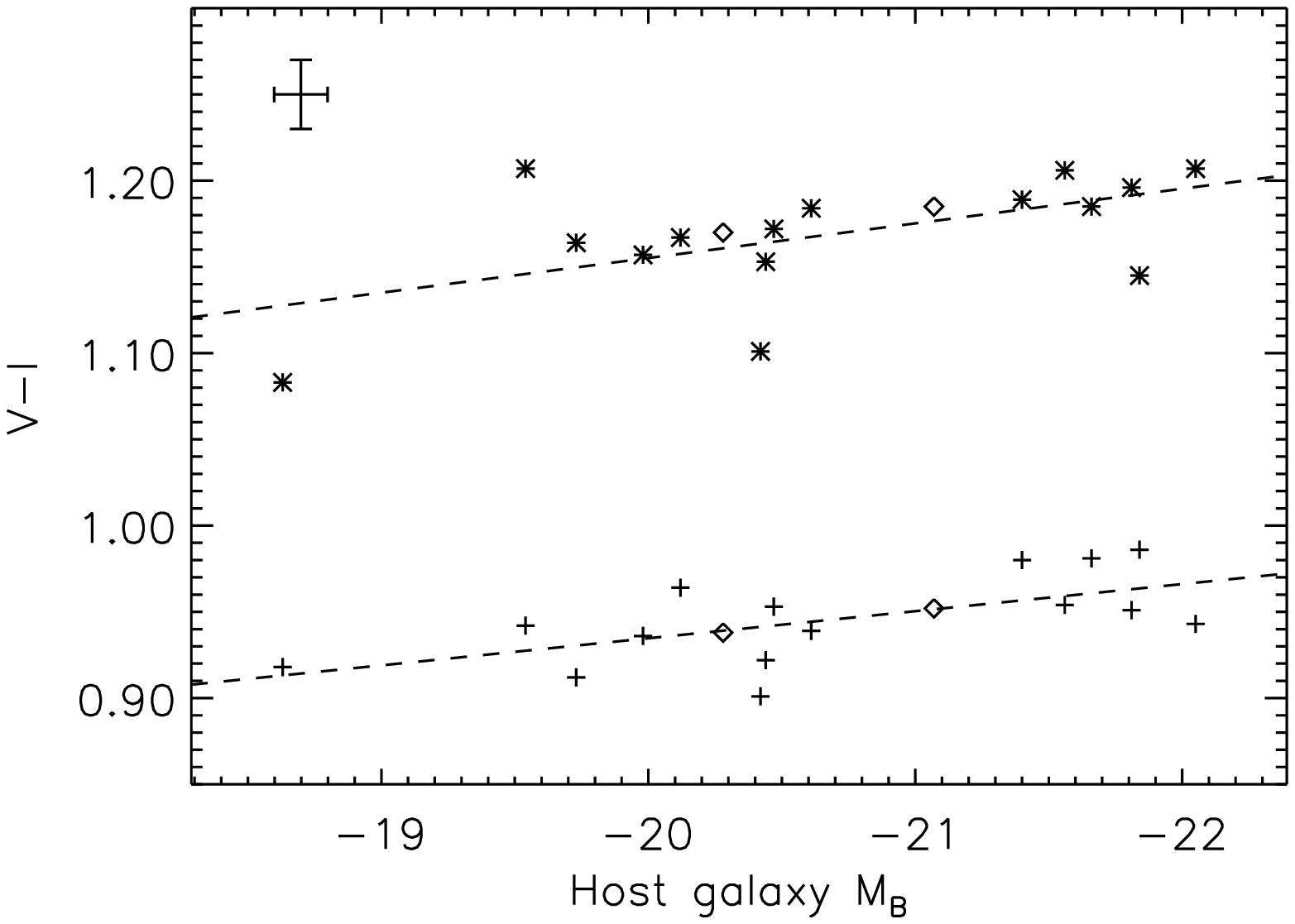}
\end{minipage}
\begin{minipage}{6.9cm}
\epsfxsize=6.9cm
\epsfbox[85 370 535 695]{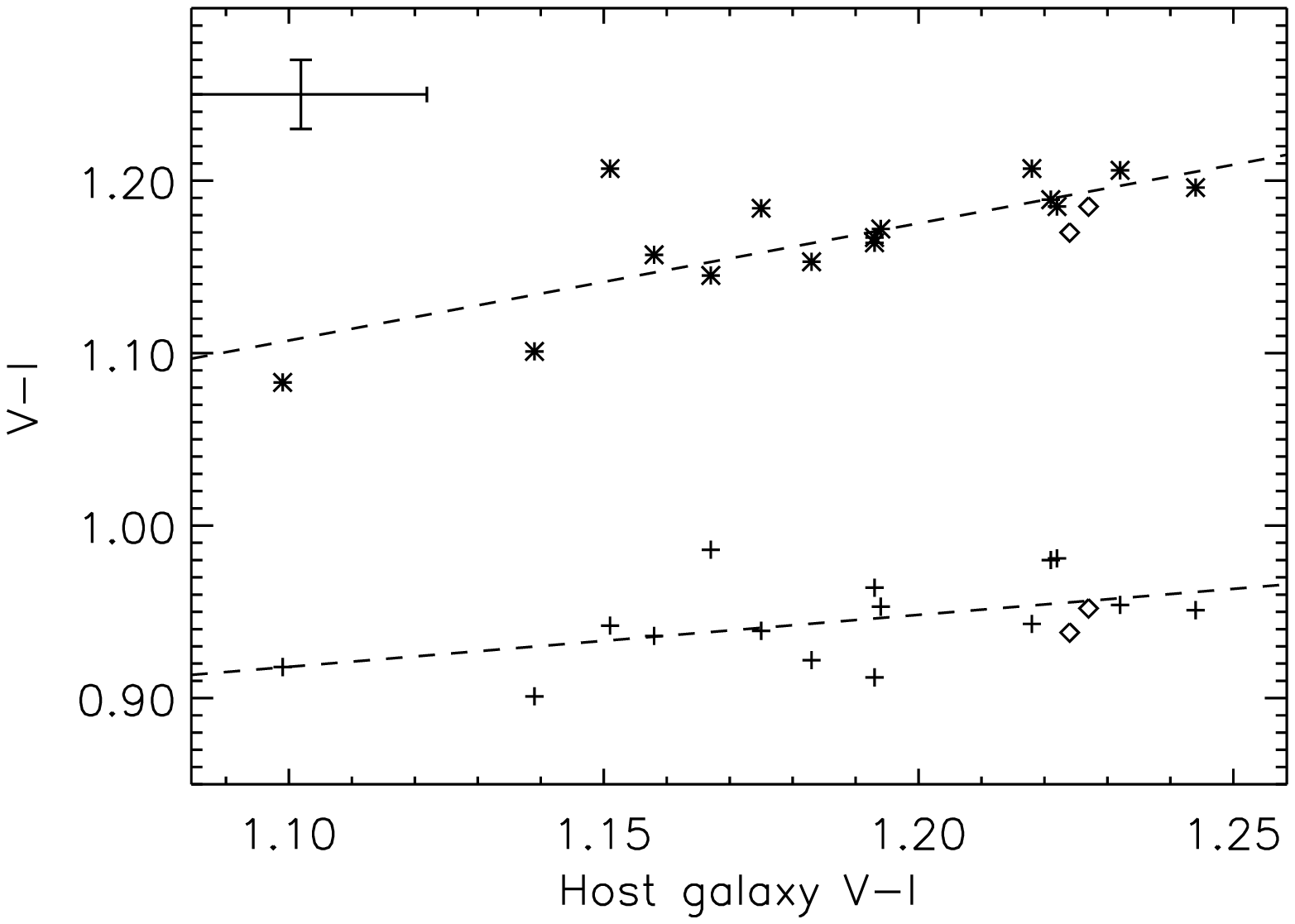}
\end{minipage}
\end{minipage}
\caption{GC peak colors versus host galaxy properties.}
\end{figure}

These results are somewhat different from previous findings (Forbes et
al.~1997; Burgarella, Kissler-Patig \& Buat 2001; Forbes \& Forte 2001) in
which the red GC colors were found to correlate strongly with parent galaxy
properties while the blue ones were thought not to correlate
significantly. This may be because of the heterogeneous nature of the data
used in these studies which tends to mask more subtle correlations.

The implication of a correlation between the properties of the blue GCs and
the parent galaxy is profound. At the time of formation, the blue GCs, or at least a significant fraction of them, must
have known about the final galaxy to which they would belong. This may present
problems for mergers and/or accretion as dominant
mechanisms in the formation history of GC systems.

\section{More Distant Galaxies}

Work is underway on the GCLFs and color distributions of a sample of more
distant galaxies. These systems are too far away for accurate size
estimates for individual clusters.

\begin{table}
\caption{\label{tab:hstpt}
  Color Distributions of Galaxies Beyond Virgo.
}
\begin{tabular}{lcccc} \hline
Galaxy & Location & Dist. Mod & $(V-I)_{0B}$  & $(V-I)_{0R}$ \\ \hline
NGC 3311 & Hydra & 33.5 & 0.91 & 1.09 \\
IC 4051 & Coma & 35 & 0.95 & 1.15 \\
NGC 4881 & Coma & 35 & 0.95 &  -  \\
NGC 5846 & Group & 32.3 & 0.96 & 1.17 \\ 
NGC 7562 & GH166 & 33 & 0.97 & - \\
NGC 7619 & Peg I & 33.5 & 0.99 & 1.24 \\ \hline \hline
\end{tabular}
\end{table}

In Table 1 we summarize our results for the peak colors of the blue and red
subpopulations. Notice that these galaxies have peak colors that are
typical of those found in the nearby sample.  NGC 3311 was previously
thought to have an almost exclusively metal-rich GC population (Secker et
al.~1995) from ground-based photometry. This would present challenges to
both the {\it in situ} and merger scenarios. However, Brodie, Larsen \&
Kissler-Patig (2000) recently showed that this galaxy has an entirely
normal color distribution based on our deep HST data. NGC 4881 appears to
be the analog of the nearby galaxy, NGC 4365, in having a single
(broadened) metal poor peak, as noted by Baum et al.~1995.  Too few clusters are assigned to the red peak
in NGC 7562 to allow a meaningful estimate of $(V-I)_{0R}$.

\section{Formation Scenarios}

As we have seen, any successful formation scenario needs to explain a very
wide range of properties of GC color distributions: 

\noindent
$\bullet$ Bimodal with roughly equal $N_{Blue}$ and $N_{Red}$ (NGC 1404, 4649,
4472).

\noindent
$\bullet$ Bimodal with much reduced $N_{Red}$ (NGC 4406).

\noindent
$\bullet$ Single (but broadened) metal poor peak (NGC 4365, 4881).

\noindent
$\bullet$ \parbox[t]{4.7in}
{Continuous color distribution spanning a range similar to true
bimodals (NGC 524, 4552).}
 
In addition, a successful model will have to accommodate the increasing
evidence that both subpopulations are old. That this is the case is borne
out by our studies of the GCLFs of the blue and red subpopulations in that
the turnover magnitudes differ only by the amount predicted by the SSP
models under the assumption of similar old ages (see Puzia et al.~1999 for
a detailed discussion). Evidence for old ages comes most convincingly from
spectroscopy but few galaxies GC systems have so far been studied with
large enough samples and adequate signal-to-noise to place interesting
constraints on age differences.  Only NGC 1399 (Kissler-Patig et al.~1998),
NGC 4472 (Beasley et al.~2000), M87 (Cohen, Blakeslee \& Ryzhov 1998), M81
(Schroder et al.~2001) and M31 (Huchra, Brodie \& Kent 1991; Perrett et al.~2001) have good enough
spectra for relative age constraints. It is likely to prove extremely
diffiult to differentiate ages greater than $\sim$8--10 Gyr, even with
superb signal-to-noise data because age contours are so closely spaced
(they even cross in some models) in index-index planes generated from SSP
models. Similar old ages would be at odds with the standard merger picture
but would be compatible
with mergers occurring at high redshift where the similarities between the
models outweigh their differences.

Particularly interesting constraints to emerge from this work are the
correlations between GC colors and host galaxy luminosity (mass) and color
for {\it both} red and blue GCs. If the correlation for the blue GCs is
verified it will be difficult to explain under both the merger and
accretion scenarios. Note too that we consistently find roughly equal
numbers of red and blue GCs, or a preponderance of blues. This is also a
challenge for these scenarios. In particular, in the merger picture, the
lower specific frequency of spirals must be increased to the higher (by a
factor of $\sim$3 for giant E galaxies) value in ellipticals by the
production (in the merger) of numerous red clusters. It is hard to escape the
conclusion that the resulting galaxy would have a preponderance of red
clusters. The accretion model is challenged only because such a large
number of blue clusters must be acquired along with minimal amounts of
starlight.

\section{What's Next?}

We do not yet understand the origin of the size difference between red and
blue GCs, whether it is primordial or the result of dynamical evolution.
To produce it by dynamical effects would require constancy in the ratio of
the perigalactic distances of the blue and red subpopulations over a wide
range of parent galaxy luminosities and environments. Were red GCs formed
on predominantly radial orbits and blue GCs formed on predominantly
circular orbits? It might be more natural to suppose that red GCs formed
from denser protocluster clumps, perhaps under conditions influenced by
their higher metallicities.

An answer to the size question will naturally require kinematic information
for large samples of GCs. This information will also inevitably assist in
differentiating between models for the origins of GC subpopulations but it
is not yet clear how to interpret such data. We are accumulating relevant
information at an increasing pace from spectroscopic programs on 8--10m
class telescopes. What is largely lacking is a theoretical basis for
comparison between the scenarios (but see Bridges article in these
proceedings for a comparison of current data with merger ideas).

Spectroscopic data also offer the best chance of age discrimination and we
can expect additional insights into the timescales for GC formation from
studies of individual element abundance ratios, in particular the ratios of
$\alpha$ elements to Fe which reflect the relative contributions of SN
Types I and II.

Much is still to be learned from photometry, particularly with
HST. Outstanding questions relevant to the work described here include: How
common are the ``faint fuzzies''?  Are they always associated with the disks of
lenticulars? What else will we find when we study the faint wings of GCLFs?
Can we see the signatures of dynamical evolution by observing galaxies with
a range of ages?

We are beginning to learn that there are many similarities in the
characteristics of the subpopulations in spirals and ellipticals. Since it
is not generally thought that spirals themselves form from {\it major} mergers
of disk systems, this might present another challenge to the standard
merger picture. However, more spiral GC systems (especially those of later
type galaxies) need to be studied before we will know how widespread these
similarities are and to what extent they will influence our ideas on GC and
galaxy formation.

\section{Conclusions}

Overall then, our results seem to fit better with an {\it in situ} scenario
in which both GC populations ``knew'' about the size of the final galaxy to
which they would belong. This implies that the initial phase of GC
formation in gE galaxies must have taken place after they assembled into
individual entities, i.e., both GC populations formed within the potential
well of the protogalaxy in multiple episodes of star formation. If the {\it in
situ} idea is right, accurate age-dating of the red GCs will pinpoint the
epoch at which the bulk of the galaxy was formed. Our best estimates
suggest this occurred $>$ 8--10 Gyr ago but it is of critical importance to
improve this constraint for comparison with models (such as hierarchical
clustering) for the formation of structure in the early universe.

\acknowledgements

I am grateful to my many collaborators, especially S{\o}ren Larsen, and to
Mike Fall for interesting discussions. This work was supported by National
Science Foundation grant number AST9900732.

\end{document}